\documentstyle[preprint,aps,eqsecnum]{revtex}

\def\beq#1{\begin{equation} \label{#1}}
\def\eeq{\end{equation}}
\def\bra#1{\left\langle #1\right\vert}
\def\ket#1{\left\vert #1\right\rangle}

\input prepictex
\input pictex
\input postpictex
\newdimen\tdim
\tdim=\unitlength
\def\stpltsmbl{\setplotsymbol ({\small .})}

\newbox\phru
\setbox\phru=\hbox{\beginpicture
\setcoordinatesystem units <\tdim,\tdim>
\stpltsmbl
\setquadratic
\plot
0 0
2.5 3
5 0
7.5 -3
10 0
/
\endpicture}
\def\photonru #1 #2 *#3 /{\multiput {\copy\phru}  at
#1 #2 *#3 10 0 /}

\newbox\sru
\setbox\sru=\hbox{\beginpicture
\setcoordinatesystem units <\tdim,\tdim>
\stpltsmbl
\setquadratic
\plot
  0.0   0.0
  4.8   1.5
  7.5   5.0
  7.3   8.5
  5.0  10.0
  2.7   8.5
  2.5   5.0
  5.2   1.5
 10.0   0.0
/
\endpicture}
\def\springru #1 #2 *#3 /{\multiput {\copy\sru}  at
#1 #2 *#3 10 0 /}

\def\PRL{ Phys. Rev. Lett.}

\begin{document}
{
\tighten

\title{FSI Rescattering in $B^\pm$ Decays via States with $\eta$, $\eta'$
$\omega$ and $\phi$}

\author{Harry J. Lipkin\,$^{a,b}$}

\address{ \vbox{\vskip 0.truecm}
  $^a\;$Department of Particle Physics \\
  Weizmann Institute of Science, Rehovot 76100, Israel \\
\vbox{\vskip 0.truecm}
$^b\;$School of Physics and Astronomy \\
Raymond and Beverly Sackler Faculty of Exact Sciences \\
Tel Aviv University, Tel Aviv, Israel\\
~\\FTLIPKIN@wiswic.weizmann.ac.il
\\~\\
}

\maketitle

\begin{abstract}

New results going beyond those obtained from isospin and flavor symmetry
and subject to clear experimental tests are obtained for effects of FSI in
$B^\pm$ decays to charmless strange final states containing neutral
flavor-mixed mesons like $\omega$, $\phi$, $\eta$ and $\eta'$. The most general
strong-interaction diagrams containing arbitrary numbers of quarks and gluons
are included with the assumptions that any $q \bar q$ pair created by gluons
must be a flavor singlet, and that there are no hairpin diagrams in which a
final meson contains a $q \bar q$ pair from the same gluon vertex. The
smallness of $K^- \eta$ suggests that it might have a large CP violation.
A sum rule is derived to test whether the large $K^- \eta'$ requires the
addition of an additional glueball or charm admixture.
Further analysis from $D_s$ decay systematics supports this picture of FSI and
raises questions about charm admixture in the $\eta'$.
\end{abstract}
} 

\renewcommand{\baselinestretch}{1.2}
\setlength{\baselineskip}{16pt}
\vspace{2ex}
The recent observation of large branching ratios for B decays into
final states containing the $\eta'$ suggests including these states
in treatments of final state rescattering. Mixtures with SU(3) singlet
and octet components are not easily treated in SU(3); thus treatments of
final state rescattering\cite{Falk} tend to omit the $K \eta'$
intermediate state. We show how to extend the standard isospin\cite{PBPENG} and
SU(3)\cite{ROSGRO,etap} treatments of $B$ decays to include flavor-mixed final
states containing $\omega$, $\phi$, $\eta$ and $\eta'$ mesons without
introducing additional free parameters. We also apply our new method to
otherwise unexplained $D_s$ decay systematics\cite{PHAWAII,CLEODS}.

We exploit known\cite{HJLCharm} flavor-topology\cite{CLOLIP12}
characteristics of quasi-two-body charmless strange decays of $B^-$ mesons.
The final states all have the quark composition $s \bar u q \bar q$
where $q \bar q$ denotes a pair of the same flavor which can be $u \bar u$ ,
$d \bar d$ or $s \bar s$, and we do not consider the possibility of charm
admixture in the final state. The $q \bar q$ pair may come from a very
complicated diagram involving many quarks and gluons. But all
possibilities for its origin are illustrated in the diagrams of figures 1-4
and have been discussed in detail\cite{CLOLIP12}. We neglect the contribution
of the electro-weak penguin diagram in this work.

Our treatment is based on the following three assumptions:

1. We neglect the contributions of ``hairpin diagrams" (see figure 1).

2. We assume that all quark-antiquark pairs created by gluons are flavor
singlets and that SU(3) flavor symmetry holds for the fragmentation process
in which the final quark-antiquark pairs create the final mesons.

3. We assume a standard pseudoscalar mixing\cite{Bramon,PengSU3},
$$ \ket{\eta} = {1\over{\sqrt 3}}\cdot(\ket{P_u} + \ket{P_d} -\ket{P_s});
 ~ ~ ~ \ket{\eta'} = {1\over{\sqrt 6}}\cdot(\ket{P_u} + \ket{P_d} +
2\ket{P_s})
 \eqno (1a)$$
$$ \ket{\pi^o} = {1\over{\sqrt 2}}\cdot(\ket{P_u} - \ket{P_d}).
  \eqno (1b)$$
where $P_u$, $P_d$ and $P_s$ denote the $u \bar u$, $d \bar d$ and $s \bar s$
components in the $\pi^o$, $\eta$ and $\eta'$ pseudoscalar mesons.

The neglect of the hairpin diagrams is based on the
Alexander-Frankfurt-Harari-Iizuka-Levin-Okubo-Rosner-Scheck-Veneziano-Zweig
rule\cite{ALS,LevFran,HarDD,Venez}, often abbreviated A...Z or OZI. Its first
controversial prediction\cite{ALS}
$\sigma(K^-p \rightarrow \Lambda \rho^o)=
\sigma(K^-p \rightarrow \Lambda \omega)$ related final states in completely
different isospin and flavor-SU(3) multiplets.
The experimental confirmation of this prediction\cite{ZGS} showed that final
state interactions did not disturb the equality between the production of two
completely different states unrelated by any known symmetry.
This OZI or A...Z rule arises in the duality diagrams\cite{HarDD} of
old-fashioned Regge phenomenology in which the leading Regge t-channel
exchanges are dual to s-channel resonances\cite{JENSEN} and in the more
modern planar quark diagrams in large $N_c$ QCD\cite{PAQMREV}.
Although no rigorous symmetry derivation has yet been found,
it has been repeatedly confirmed in a large variety of experimental results
and theoretical analyses for strong interaction three-point and four-point
functions\cite{ALS,Venez,Exter}.

We therefore neglect the hairpin diagram to obtain predictive power which can
be tested with future experimental data. Our first prediction is
$$ BR (B^\pm \rightarrow K^\pm \omega) = BR (B^\pm \rightarrow K^\pm \rho^o)
\eqno(2)                                          $$
because the $\rho^o$ and $\omega$ mesons both come only from the
$\overline u u$ quark state in all diagrams described by figures 2-4.
A previous derivation\cite{PENGRHO} justified the escape of the final mesons
without flavor change by a hand-waving asymptotic freedom argument.
The diagrams of figures 2-4 show here that relations like (2) require only
exclusion of hairpin diagrams and hold even in the presence of strong final
state rescattering via all other quark-gluon diagrams.

We now note that there are only two possible mechanisms for the creation of the
$q \bar q$ pair in the final states $s \bar u q \bar q$ created by the diagrams
of figures 2-4.

(1) It is created by gluons and must therefore be a flavor singlet denoted by
$(\bar q q)_1$ (see figure 2); The transition to a final two-pseudoscalar state
is therefore:
$$ s G \bar u \rightarrow  s (\bar q q)_1 \bar u
\rightarrow {1\over{\sqrt 3}}\cdot(\ket{K^- P_u} + \ket{\bar K^o\pi^-}
+ \ket{P_s K^-})\equiv  \ket{R_{PP}} \eqno (3a)$$
The state $\ket{R_{PP}}$ defined as the final two-pseudoscalar state
produced by the strong pair creation diagram can be rewritten
 $$\ket{R_{PP}} = {1\over{\sqrt 6}}\cdot \ket{K^- \pi^o} +
{1\over{\sqrt 2}}\cdot \ket{K^- \eta' } +
{\xi\over{\sqrt 2}}\cdot \ket{K^- \eta } +
{1\over{\sqrt 3}}\cdot\ket{\bar K^o\pi^-}
 \eqno (3b)$$
where $\xi$ is a small parameter to introduce a $K \eta$ contribution which
vanishes in the SU(3) symmetry limit with the particular mixing\cite{Bramon}
angle (1) as a result of a cancellation between the contributions from the
$P_u$ and $P_s$ components in the $\eta$ wave function. A small but finite
value of $\xi$ is suggested for realistic models by the $K \eta$ suppression
observed in other experimental transitions\cite{HJLCharm,PKETA} like decays of
strong $K^*$ resonances known to proceed via an even parity $u \bar  s$
+ singlet state. The possibility of a relatively large $CP$ violation in a
small $K \eta$ branching ratio is discussed below.

(2) The quark is a $u$ quark from the weak
vertex and the pair can only be $u \bar u$ (see figures 3 \& 4).
The transition to a final two-pseudoscalar state is therefore:
$$  s \bar u u \bar u\rightarrow  \ket{K^- P_u}
\eqno (4)$$
Only the coherent sum of the amplitudes from the color-favored (see figure 3)
and color-suppressed (see figure 4) diagrams is relevant for these charged
decays. This simplification provides predictive power and allows crucial tests
of the basic assumptions, but does not arise in neutral decays where the two
antiquarks have different flavors and the two diagrams lead to different
final states.

The decays are thus described by three parameters, an $\ket{R_{PP}}$ amplitude
produced by the diagram of figure 2, a $\ket{K^- P_u} $ amplitude
produced by the sum of the contributions from the diagrams of figures 3 and 4,
and a relative phase. The one relation obtainable between the
decays to four final states is shown below to be the sum rule:
$$ \tilde \Gamma(B^\pm \rightarrow K^\pm \eta') + \tilde \Gamma(B^\pm
\rightarrow K^\pm \eta) = \tilde \Gamma(B^\pm \rightarrow K^\pm \pi^o)
+  \tilde \Gamma(B^\pm \rightarrow \tilde K^o \pi^\pm)
\eqno(5a)                                          $$
where $\tilde \Gamma$ denotes the theoretical partial width without phase
space corrections. $\tilde K^o$ denotes $ K^o$ for the $B^+$ decay and
$\bar K^o$ for the $B^-$ decay. The sum rule is independent of the
$\eta - \eta'$ mixing angle.

These sum rules are of particular interest because of the large experimentally
observed branching ratio\cite{CLEO} for $B^\pm \rightarrow K^\pm \eta'$. A
violation favoring $B^\pm \rightarrow K^\pm \eta'$ can provide convincing
evidence for an additional contribution\cite{PengSU3} like a glueball, charm
admixture\cite{Hareta,FRIJACK,HALZHIT,ATSON} in the $\eta'$ wave function or an
A...Z-violating hairpin diagram. Present data are not statistically
significant.
If better data show a violation, the best candidate seems to be the charm
admixture originally suggested\cite{Hareta} to explain the anomalously large
A...Z-violating cascade decay $\psi' \rightarrow \eta \psi$. An A...Z-violating
gluonic hairpin like the one shown in figure 1 would also contribute to the
analogous cascade $\Upsilon(nS) \rightarrow \eta \Upsilon(1S)$ which
so far has not been seen\cite{PDG}.

In the kaon-vector (KV) system the ideal mixing of the $\omega-\phi$ system
simplifies the treatment to give two equalities; namely eq. (2) and
$$ \tilde \Gamma(B^\pm \rightarrow K^\pm \phi)
= \tilde \Gamma(B^\pm \rightarrow K^o \rho^\pm)
\eqno(5b)                                          $$
Both the sum rule (5a) and the equality (5b) assume SU(3) flavor symmetry
for the diagram of figure 2.

The relations (2) and (5b) should provide good experimental tests of our
basic assumptions before the data are precise enough to check CP
violation. The relation (2) tests the assumption that only the diagrams
shown in figures (2-4) contribute. The relations (5b) also require the
SU(3) relation between strange and nonstrange pair production in the diagram
of figure 2. The experimental test of (5b) can provide useful input to
estimate SU(3) symmetry breaking in the sum rule (5a), particularly if an
experimental disagreement suggests an additional component in the $\eta'$ wave
function.

We now examine the dependence of these amplitudes on CKM matrix elements. The
two $b$ quark weak decay vertices contributing to these decays are:
  $$b \rightarrow u \bar u s                    \eqno(6a) $$
  $$b \rightarrow Q \bar Q s                   \eqno(6b) $$
where Q is a heavy quark, $c$ or $t$. The two vertices depend upon
two different products of CKM matrix elements. Their interference can give rise
to direct CP violation.

The dominant contribution to the charmless strange B decays is now generally
believed to arise from the ``gluonic penguin" diagram; the next from the tree
diagram. We avoid controversies about what exactly is a penguin and how to
include final state interactions by defining amplitudes in terms of their quark
decay vertices (6a) or (6b) and flavor topology diagrams (figures 2-4).
There are three possible amplitudes which we denote by $A$, $B$ and $C$.

A: Quark vertex $b \rightarrow Q \bar Q s$;  Strong pair creation diagram
(figure 2). This gives
$$ B^-(\bar u b) \rightarrow \bar u Q \bar Q s \rightarrow \bar u
(q \bar q)_1 s \rightarrow \ket{R_{PP}} \eqno (7a)$$
B: Quark vertex $b \rightarrow u \bar u s$;  Sum of weak pair creation diagrams
(figures 3-4). This gives
$$ B^-(\bar u b)  \rightarrow \bar u u \bar u s \rightarrow \ket{K^- P_u}
\eqno (7b)$$

C: Quark vertex $b \rightarrow u \bar u s$;  Strong pair creation diagram
(figure 2). This gives
$$ B^-(\bar u b)  \rightarrow \bar u u \bar u s \rightarrow \bar u
(q \bar q)_1 s \rightarrow \ket{R_{PP}}
\eqno (7c)$$

The $A$ amplitude includes not only the ``normal" dominant penguin
diagram\cite{PENGRHO,PKETA} but also other diagrams proportional to the
$b \rightarrow Q \bar Q s$ vertex where the $Q \bar Q $ pair is annihilated via
a final state interaction and sums over contributions from both $c \bar c $
and $t \bar t$ pairs.
The $B$ amplitude is the tree including flavor-conserving final state
interactions, and the $C$ amplitude is a tree followed by a flavor-changing
final state interaction. These are
the only possibilities, if the gluonic hairpin diagram is
excluded, and include all possible final state interactions that can take
place in the black boxes of figures (2-4).

The relative magnitudes and strong phases of these amplitudes are model
dependent. They are simply related to the isospin and SU(3) amplitudes
conventionally used to treat $B \rightarrow K \pi$ decays and give no new
information for these analyses\cite{PBPENG,ROSGRO}. The new ingredient
introduced by flavor-topology analyses\cite{CLOLIP12} is the inclusion of the
neutral flavor-mixed meson states. This allows the extension to the $K \eta$
and $K \eta'$ modes of any dynamical or phenomenological treatment of
$B \rightarrow K \pi$ decays\cite{Falk}without introducing additional
parameters.

The transition matrix for a $B^-$ decay into any kaon-pseudoscalar state
$\ket{f}$ can be written as the sum of three terms proportional to the
three amplitudes $A$, $B$ and $C$.
$$ \bra{f} T \ket{B^-} = (A + C )\langle f \ket{R_{PP}}
+ B \langle f \ket{K^- P_u}
 \eqno(8)$$
Note that the amplitudes $A$ and $C$ contribute only via the sum $A + C$ since
both lead to the same final state $\ket R$.

Substituting the relations (3b) into (8) then gives the relations
$$ \bra{\bar K^o\pi^-} T \ket{B^-} = {{A + C} \over{\sqrt 3}}; ~ ~ ~
\bra{K^- \pi^o} T \ket{B^-} = {{A + C} \over{\sqrt 6}} + {B \over{\sqrt 2}}
\eqno(9a)$$
$$ \bra{K^-\eta} T \ket{B^-} = {B\over{\sqrt 3}} +
{{\xi (A + C)} \over{\sqrt 2}}
; ~ ~ ~ \bra{K^-\eta'} T \ket{B^-} = {{A + C} \over{\sqrt 2}} +
{B \over{\sqrt 6}}   \eqno(9b)$$
$$ \tilde \Gamma(B^- \rightarrow \bar K^o\pi^-) =
{{|A + C|^2 } \over{3}}
 \eqno(10a)$$
$$ \tilde \Gamma(B^- \rightarrow K^- \pi^o) =
 {{|A + C|^2 } \over{6}} + {|B|^2\over{2}}
+ {{|A + C||B|\cos \theta} \over{\sqrt 3}}
 \eqno(10b)$$
$$ \tilde \Gamma(B^- \rightarrow K^- \eta) =  {|B|^2\over{3}} +
{{\xi^2 (|A + C|^2 ) } \over{2}} +
{{2 \xi |A + C||B|\cos \theta} \over{\sqrt 6}}
 \eqno(10c)$$
$$ \tilde \Gamma(B^-\rightarrow K^- \eta') = {{|A + C|^2 } \over{2}}
+ {|B|^2\over{6}} + {{|A + C||B|\cos \theta} \over{\sqrt 3}}
\eqno(10d)$$
where $\theta$ denotes the relative phase of the $B$ amplitude and the sum
of the $A$ and $C$ amplitudes.

Direct CP-violation asymmetries are obtained from interference between the $A$
amplitude and the $B$ and $C$ amplitudes which have different weak phases.
In the standard model the amplitudes for charge conjugate transitions have the
same magnitude and the same strong phase but have opposite weak phase. Thus
the transition matrix for a $B^+$ decay into any kaon-pseudoscalar state
$\ket{\bar f}$ which is the charge conjugate of the state $\ket{f}$ in eq.
(8) can be written as exactly the same expression (8) in terms of the
same three amplitudes $A$, $B$ and $C$, except for modified weak phase factors.

This is most conveniently displayed by writing the amplitudes explicitly in
terms of their weak and strong phases,
$$ A \equiv |A|e^{iS_A} e^{iW_A}; ~ ~ ~  B \equiv |B|e^{iS_B} e^{iW_B} ; ~ ~ ~
C \equiv |C|e^{iS_C} e^{iW_B} \eqno(11a)   $$
where $S_A$, $S_B$ and $S_C$ denote the strong phases of the amplitudes
$A$, $B$ and $C$ and similarly for the weak phases, and we have noted that
$W_B = W_C$ since they depend upon the same CKM matrix elements.
It is also convenient to define three relevant phase differences
$$ \phi_{SB} \equiv S_A - S_B; ~ ~ ~  \phi_{SC} \equiv S_A - S_C; ~ ~ ~
\phi_w \equiv W_A - W_B \eqno(11b)   $$
We can then rewrite eq.(8)
$$ e^{-iS_A} e^{-iW_A} \bra{f} T \ket{B^-} = (A + Ce^{-i\phi_{SC}}
e^{-i\phi_w})\langle f \ket{R_{PP}} + Be^{-i\phi_{SB}} e^{-i\phi_w}
\langle f \ket{K^- P_u}  \eqno(12a)$$
where A, B, and C now denote the magnitudes of these amplitudes.
The charge conjugate amplitude and the direct CP violation asymmetries
can then be written.
$$ e^{-iS_A} e^{iW_A} \bra{\bar f} T \ket{B^+} = (A + Ce^{-i\phi_{SC}}
e^{i\phi_w})\langle f \ket{R_{PP}}+ Be^{-i\phi_{SB}}e^{i\phi_w}
\langle f \ket{K^- P_u}  \eqno(12b)$$
$$ |\bra{\bar f} T \ket{B^+}|^2 - |\bra{f} T \ket{B^-}|^2 =
4A \sin \phi_w \{C \sin \phi_{SC} \langle f \ket{R_{PP}} + B \sin \phi_{SB}
\langle f \ket{K^- P_u}\}
 \eqno(13)$$

$$ \tilde \Gamma(B^+ \rightarrow K^o\pi^+) -
\tilde \Gamma(B^- \rightarrow \bar K^o\pi^-) =
{{4AC \sin \phi_w \sin \phi_{SC}} \over{3}}
\eqno(14a)$$
$$
\tilde \Gamma(B^+ \rightarrow K^+ \pi^o) -
\tilde \Gamma(B^- \rightarrow K^- \pi^o)  =
 {{4AC \sin \phi_w \sin \phi_{SC}} \over{6}} +
{{2AB \sin \phi_w \sin \phi_{SB}} \over{\sqrt 3}}
  \eqno(14b)$$
$$ \tilde \Gamma(B^+ \rightarrow K^+ \eta ) -
\tilde \Gamma(B^- \rightarrow K^- \eta)  =
{{4 \xi AB \sin \phi_w \sin \phi_{SB}} \over{\sqrt 6}}
 \eqno(14c)$$
$$ \tilde \Gamma(B^+\rightarrow K^+ \eta') -
\tilde \Gamma(B^-\rightarrow K^- \eta')
 = {{4AC \sin \phi_w \sin \phi_{SC}}
\over{2}} + {{2AB \sin \phi_w \sin \phi_{SB}} \over{\sqrt 3}}
\eqno(14d)$$

The $A$ amplitude is dominated by the penguin and expected to be much larger
than the $B$ and $C$ amplitudes. Thus $\Gamma(B^- \rightarrow K^- \eta)$ is
expected to be much smaller than for the other decays. However, to first order
in the small parameter $\xi$ and the small ratios $B/A$ and $C/A$,
  $$ {{\tilde \Gamma(B^+ \rightarrow K^+ \eta) -
\tilde \Gamma(B^- \rightarrow K^- \eta )}
\over
{\tilde \Gamma(B^+ \rightarrow K^+ \eta) +
\tilde \Gamma(B^- \rightarrow K^- \eta )}} \approx
{{6 \xi AB \sin \phi_w \sin \phi_{SB}} \over{\sqrt 6 B^2}}
 \eqno(15)$$
This is of order $(\xi A/B)$ while the analogous relative
asymmetries for other decay modes are of order $(B/A)$ and $(C/A)$. Thus even
though the signal for $CP$ violation (14c) may be small for
$B^+ \rightarrow K^+ \eta$, the signal/background ratio (15) may be more
favorable. An exact theoretical calculation of $\xi$ is not feasible. A good
estimate from future data may enable a choice  between different
decay modes as candidates for observation of direct CP violation.

Higher resonances can be incorporated into the final state rescattering with
simplifications from $C$, $P$, Bose symmetry and flavor $SU(3)$. Since
the vector-pseudoscalar states have opposite parity, the next higher
quasi-two-body final states allowed by conservation laws are the vector-vector
s-wave and d-wave states. These can be incorporated by using models for the
decays of a scalar resonance into these channels and inputs from polarization
measurements and branching ratios for the vector-vector states.

The same approach can be used to treat vector-pseudoscalar final states.
Expressions for the $K\rho$ decay modes are obtained directly from eqs. (9)
and (14) for the $K\pi$ modes and the $K\omega$ and $K\phi$ decays are
given by eqs. (2).

For the $K^*P$ system the analogs of eqs. (3), (8),(10) and (14) are
 $$\ket{R_{VP}} = {1\over{\sqrt 3}}\cdot(\ket{K^{*-} P_u} +
\ket{\bar K^{*o}\pi^-} + \ket{P_s K^{*-}}) =
{1\over{\sqrt 3}}\cdot(\ket{K^{*-} P_u} + \ket{\bar K^{*o}\pi^-}
- \ket{K^{*-} P_s})
 \eqno (16a) $$
 $$\ket{R_{VP}} = {1\over{\sqrt 6}}\cdot \ket{K^{*-} \pi^o} -
{1\over{3 \sqrt 2}}\cdot \ket{K^{*-} \eta' } +
{2\over{3}}\cdot \ket{K^{*-} \eta } +
{1\over{\sqrt 3}}\cdot\ket{\bar K^{*o}\pi^-}
 \eqno (16b)$$
$$ \bra{f} T \ket{B^-} = (A_{VP} + C_{VP} )\langle f \ket{R_{PP}} +
B_{VP} \langle f \ket{K^{*-} P_u}
 \eqno(17)$$
$$ \tilde \Gamma(B^- \rightarrow \bar K^{*o}\pi^-) =
{{|A_{VP} + C_{VP}|^2 } \over{3}}
\eqno(18a)$$
$$ \tilde \Gamma(B^- \rightarrow K^{*-} \pi^o) =
 {{|A_{VP} + C_{VP}|^2 } \over{6}} + {|B_{VP}|^2\over{2}}
+ {{|A_{VP} + C_{VP}||B_{VP}|\cos \theta_{VP}} \over{\sqrt 3}}
  \eqno(18b)$$
$$ \tilde \Gamma(B^- \rightarrow K^{*-} \eta) =  {B_{VP}^2\over{3}} +
{{4 |A_{VP} + C_{VP}|^2 } \over{9}} +
{{4 |A_{VP} + C_{VP}||B_{VP}|\cos \theta_{VP}}   \over{3 \sqrt 3}}
 \eqno(18c)$$
$$ \tilde \Gamma(B^-\rightarrow K^{*-} \eta') =
{{|A_{VP} + C_{VP}|^2 } \over{18}} +
{B_{VP}^2\over{6}} -
{{|A_{VP} + C_{VP}||B_{VP}|\cos \theta_{VP}} \over{3 \sqrt 3}}
\eqno(18d)$$
$$ \tilde \Gamma(B^\pm \rightarrow K^{*\pm} \eta') + \tilde \Gamma(B^\pm
\rightarrow K^{*\pm} \eta) =  \tilde \Gamma(B^\pm \rightarrow K^{*\pm} \pi^o)
+  \tilde \Gamma(B^\pm \rightarrow \tilde K^{*o} \pi^\pm)
\eqno(19)                                          $$
$$  \tilde \Gamma(B^+ \rightarrow K^{*o}\pi^+) -
\tilde \Gamma(B^- \rightarrow \bar K^{*o}\pi^-) =
{{4A_{VP}C_{VP} \sin \phi_w \sin \phi_{SC}} \over{3}}
\eqno(20a)$$
$$ \tilde \Gamma(B^+ \rightarrow K^{*+} \pi^o) -
\tilde \Gamma(B^- \rightarrow K^{*-} \pi^o) =
 {{4A_{VP}C_{VP} \sin \phi_w \sin \phi_{SC}} \over{6}} +
{{2A_{VP}B_{VP} \sin \phi_w \sin \phi_{SB}} \over{\sqrt 3}}
  \eqno(20b)$$
$$ \tilde \Gamma(B^+ \rightarrow K^{*+} \eta) -
\tilde \Gamma(B^- \rightarrow K^{*-} \eta ) =
{{16 A_{VP}C_{VP} \sin \phi_w \sin \phi_{SC}} \over{9}} +
{{8 A_{VP}B_{VP} \sin \phi_w \sin \phi_{SB}} \over{3\sqrt 3}}
 \eqno(20c)$$
$$ \tilde \Gamma(B^+\rightarrow K^{*+} \eta') -
\tilde \Gamma(B^-\rightarrow K^{*-} \eta') =
{{4 A_{VP}C_{VP} \sin \phi_w \sin \phi_{SC}} \over{18}}
- {{2 A_{VP}B_{VP} \sin \phi_w \sin \phi_{SB}} \over{3\sqrt 3}}
\eqno(20d)$$
where the subscript $VP$ denotes $K^*$-pseudoscalar amplitudes and phases.
The weak phase $\phi_w$ has the same value as for the two-pseudoscalar
case. The strong phases $\phi_{SC}$ and $\phi_{SB}$ are different.

The odd parity of the final state is seen in eqs. (16) to lead to a reversal
of the relative phase of the strange and nonstrange contributions of the
$\eta$ and $\eta'$. This reversal of $\eta'/\eta$ ratio has been
suggested\cite{PHAWAII,PengSU3} as a test for the presence of an
additional component in the $\eta'$ which would enhance the $\eta'$ in both
cases. We improve their quantitative prediction based only on the $A_{VP}$
amplitude with a relation which also includes the contributions from $B_{VP}$
and $C_{VP}$
$$ \tilde \Gamma(B^\pm \rightarrow K^{*\pm} \eta') =
(1/3)\cdot  \tilde \Gamma(B^\pm \rightarrow K^{*o} \pi^\pm) -
(1/3)\cdot  \tilde \Gamma(B^\pm \rightarrow K^{*\pm} \pi^o) +
{B_{VP}^2\over{3}} \eqno(21)                                          $$

The same approach can be used to treat the corresponding charmless nonstrange
decays. Simple SU(3) relations between corresponding strange and nonstrange
amplitudes can be used since replacing an s quark by a d quark in an amplitude
containing no other s or d flavors is a simple U-spin Weyl reflection. The
decays $B^\pm \rightarrow \pi^\pm \eta'$, $B^\pm \rightarrow \pi^\pm \eta$,
$B^\pm \rightarrow \rho^\pm \eta'$ and $B^\pm \rightarrow \rho^\pm \eta$ are
of particular interest since only the nonstrange components of the $\eta$ and
$\eta'$ can contribute and any enhancement of the $\eta'/\eta$ ratio is a
clear indication of an additional singlet component in the $\eta'$. If it arises
from a charmonium component in the $\eta'$ the decay is an excellent candidate
for direct CP violation.

We now note one other case which also suggests that A...Z allowed transitions
via a $q \bar q $ + singlet intermediate state may be a general feature of
final state interactions which warrants further investigation. Such transitions
can enhance the $K^*\bar K$ and $K^*\bar K^*$ modes in both $D^o$
and $D_s$ decays relative to $\phi \pi$ and $\phi \rho$. This enhances the
color-favored tree diagrams for $D^o$ decays and the color-suppressed tree
diagrams for $D_s$ decays and can explain the
dramatic change in color suppression noted\cite{PHAWAII} between the $D$ and
$D_s$ decays whose tree diagrams differ only by the flavor of a spectator
quark.

The nonstrange vector-pseudoscalar modes in both $D$ and $B$ decays already
present other puzzles\cite{PHAWAII} which surprise theorists and provide
interesting opportunities for future investigations. The role of $G$ parity
has
been pointed out with reference to the $\eta \pi$, $\eta' \pi$, $\eta \rho$,
and $\eta' \rho$ for the $D_s$ decays where four channels with
different parities and $G$ parities are not mixed by strong final
state interactions\cite{PHAWAII}. The same is also true for $B$ and $B_s$
decays. For the $VP$ decays, which have a definite odd parity, there are still
two channels. One has odd G-parity like the pion and couples to $\rho \pi$;
the
other has exotic even $G$ and couples to $\omega \pi$, $\eta \rho$, and $\eta'
\rho$. This even-G state does not couple to any $q \bar q$ state containing no
additional gluons. It therefore does not couple to any single meson
resonances,
nor to the state produced by an annihilation diagram with no gluons emitted by
the initial state before annihilation\cite{PHAWAII}. We now note that the
coupling of the even-G state is A...Z-forbidden in the present model also for
annihilation diagrams with additional gluons present because of cancellation
between contributions from the $u  \bar u$ and $d \bar d$ components of the
$\omega$, $\eta$, and $\eta'$ wave functions, whereas the two contributions
add
in the $D_s \rightarrow \rho \pi$. Here the presence of charm in the
$\eta'$ wave function may be significant because of the generally overlooked
contribution of the ``backward" weak diagram $s \rightarrow c \bar u d$.

Comparison of corresponding $D_s$ and $D$ decays into final states containing
the $\eta$ and $\eta'$ mesons have been suggested\cite{PKETA} as a means to
test for the breaking of the nonet picture by additional flavor singlet
components.

Further information which may provide important clues to this complicated
four-channel system may be obtained from angular distributions in the
$K \bar K \pi$ modes, including the $K^* \bar K$ and $K \bar K^*$ modes which
decay into $K \bar K \pi^o$. The VP $K^* \bar K$ and $K \bar K^*$ modes are
not individually eigenstates of $G$ parity and have unique p-wave angular
distributions for the vector-pseudoscalar states. The $G$-parity eigenstates
are coherent linear combinations of the two with opposite phase. They have
opposite relative parity in the $K \bar K$ system, even though
they are not produced from the same resonance. This relative parity can be
observed as constructive or destructive interference in the kinematic region
in the Dalitz plot where the two $K^*$ bands overlap. In a region where s and
p waves dominate the angular distribution of the $K \bar K$ momentum in the
$K \bar K$ center of mass system of the $K \bar K \pi$ final state relative to
the pion momentum, one $G$ eigenstate will have a $\sin^2\theta$ distribution,
the other will have a $\cos^2\theta$ distribution and interference between the
two $G$ eigenstates can show up as a forward-backward asymmetry.

A general theorem from CPT invariance\cite{PENGRHO} shows that all observed CP
asymmetries must cancel when the data are summed over all final states or over
any set of final states which are eigenstates of the strong-interaction
S-matrix. Any CP asymmetry arising in a given
channel must be canceled by an opposite CP asymmetry in some other
channels. In the case of the model described by eqs. (14), this can occur only
if there is a definite relation between the $A\cdot C$ and $A\cdot B$
interference terms. Any total CP asymmetry arising in a finite set of final
states indicates significant strong interaction rescattering between these
states and others outside the set; e.g. vector-vector or multiparticle final
states. This casts doubt on theoretical estimates of direct $CP$ violation
which do not include such rescattering.

Other attempts to estimating soft strong effects on CP violation in weak
decays\cite{Donoghue} have used Regge phenomenology with parameters obtained
from fits\cite{PDG} to total cross section data. These fits are unfortunately
highly controversial and unreliable\cite{PAQMREV}. Better fits to the same
data with completely different parameters\cite{PAQMREV,Exter} have been
obtained by using the physics input described above. A recent
Regge phenomenology calculation\cite{Falk} for $B\rightarrow K\pi$ decays
using PDG parameters\cite{PDG} shows neither a dominant effect of order unity
nor an insignificant effect of order 1\%. Further improvement seems unlikely.
In contrast the approach presented here uses well defined physics input
subject to experimental tests. If these tests are successful they can lead the
way to a considerable simplification in future treatments of FSI.

\acknowledgments
It is a pleasure to thank Edmond Berger, Karl Berkelman, Yuval Grossman,
Yosef Nir and J. G. Smith for helpful discussions and comments, and in
particular to thank Howard Georgi for extensive criticism of the previous
version. This work was partially supported by the German-Israeli Foundation
for Scientific Research and Development (GIF).

{
\tighten

}

{\begin{figure}[htb]
$$\beginpicture
\setcoordinatesystem units <\tdim,\tdim>
\stpltsmbl
\putrule from -25 -30 to 50 -30
\putrule from -25 -30 to -25 30
\putrule from -25 30 to 50 30
\putrule from 50 -30 to 50 30
\plot -25 -20 -50 -20 /
\plot -25 20 -50 20 /
\plot 50 20 120 40 /
\plot 50 0 120 20 /
\springru 50 -20 *3 /
\plot 120 -20 90 -20 120 -40 /
\put {$b$} [b] at -50 25
\put {$\overline{u}$} [t] at -50 -25
\put {$s$} [l] at 125 40
\put {$\overline{u}$} [l] at 125 20
\put {$q$} [l] at 125 -20
\put {$\overline{q}$} [l] at 125 -40
\put {$\Biggr\}$ meson} [l] at 135 30
\put {$\Biggr\}$ meson} [l] at 135 -30
\put {$G$} [t] at 70 -25
\setplotsymbol ({\large .})
\plot -15 40 60 -40 /
\plot -15 -40 60 40 /
\setshadegrid span <1.5\unitlength>
\hshade -30 -25 50 30 -25 50 /
\linethickness=0pt
\putrule from 0 0 to 0 60
\endpicture$$
\caption{\label{fig-1}} \hfill Forbidden ``gluonic hairpin'' diagram. $G$
denotes any number of gluons. \hfill~
\end{figure}}

{\begin{figure}[htb]
$$\beginpicture
\setcoordinatesystem units <\tdim,\tdim>
\stpltsmbl
\putrule from -25 -30 to 50 -30
\putrule from -25 -30 to -25 30
\putrule from -25 30 to 50 30
\putrule from 50 -30 to 50 30
\plot -25 -20 -50 -20 /
\plot -25 20 -50 20 /
\plot 50 20 120 40 /
\plot 50 -20 120 -40 /
\springru 50 0 *3 /
\plot 120 20 90 0 120 -20 /
\put {$b$} [b] at -50 25
\put {$\overline{u}$} [t] at -50 -25
\put {$s$} [l] at 125 40
\put {$\overline{q}$} [l] at 125 20
\put {$q$} [l] at 125 -20
\put {$\overline{u}$} [l] at 125 -40
\put {$\Biggr\}$ meson} [l] at 135 30
\put {$\Biggr\}$ meson} [l] at 135 -30
\put {$G$} [t] at 70 -5
\setshadegrid span <1.5\unitlength>
\hshade -30 -25 50 30 -25 50 /
\linethickness=0pt
\putrule from 0 0 to 0 60
\endpicture$$
\caption{\label{fig-2}} \hfill Strong pair creation. $G$ denotes any number of
gluons. \hfill~ \end{figure}}

{\begin{figure}[htb]
$$\beginpicture
\setcoordinatesystem units <\tdim,\tdim>
\stpltsmbl
\putrule from -25 -30 to 50 -30
\putrule from -25 -30 to -25 30
\putrule from -25 30 to 50 30
\putrule from 50 -30 to 50 30
\plot -25 -20 -50 -20 /
\plot -25 20 -50 20 /
\plot 50 0 120 -20 /
\plot 50 -20 120 -40 /
\photonru 50 20 *3 /
\plot 120 40 90 20 120 20 /
\put {$b$} [b] at -50 25
\put {$\overline{u}$} [t] at -50 -25
\put {$s$} [l] at 125 40
\put {$\overline{u}$} [l] at 125 20
\put {$u$} [l] at 125 -20
\put {$\overline{u}$} [l] at 125 -40
\put {$\Biggr\}$ meson} [l] at 135 30
\put {$\Biggr\}$ meson} [l] at 135 -30
\put {$W$} [t] at 70 15
\setshadegrid span <1.5\unitlength>
\hshade -30 -25 50 30 -25 50 /
\linethickness=0pt
\putrule from 0 0 to 0 60
\endpicture$$
\caption{\label{fig-3}} \hfill Weak pair creation. Color favored diagram.
 \hfill~ \end{figure}}

{\begin{figure}[htb]
$$\beginpicture
\setcoordinatesystem units <\tdim,\tdim>
\stpltsmbl
\putrule from -25 -30 to 50 -30
\putrule from -25 -30 to -25 30
\putrule from -25 30 to 50 30
\putrule from 50 -30 to 50 30
\plot -25 -20 -50 -20 /
\plot -25 20 -50 20 /
\plot 50 20 120 40 /
\plot 50 -20 120 -40 /
\photonru 50 0 *3 /
\plot 120 20 90 0 120 -20 /
\put {$b$} [b] at -50 25
\put {$\overline{u}$} [t] at -50 -25
\put {$u$} [l] at 125 40
\put {$\overline{u}$} [l] at 125 20
\put {$s$} [l] at 125 -20
\put {$\overline{u}$} [l] at 125 -40
\put {$\Biggr\}$ meson} [l] at 135 30
\put {$\Biggr\}$ meson} [l] at 135 -30
\put {$W$} [t] at 70 -5
\setshadegrid span <1.5\unitlength>
\hshade -30 -25 50 30 -25 50 /
\linethickness=0pt
\putrule from 0 0 to 0 60
\endpicture$$
\caption{\label{fig-4}} \hfill Weak pair creation. Color suppressed diagram.
\hfill~ \end{figure}}

\end{document}